\documentclass[aps,pre,twocolumn,showpacs,superscriptaddress]{revtex4-1}
\usepackage{graphicx}
\usepackage{dcolumn}
\usepackage{bm}
\usepackage{amssymb}
\usepackage{amsmath}
\usepackage{subfigure}
\usepackage{float}
\begin{document}
\title{Magnetization Dynamics driven by Non-equilibrium Spin-Orbit Coupled Electron Gas}
\author{Yong Wang}
\affiliation{Department of Physics, The University of Hong Kong, Hong Kong SAR, China}
\author{Wei-qiang Chen}
\affiliation{Department of Physics, South University of Science and Technology of China, China}
\author{Fu-Chun Zhang}
\affiliation{Department of Physics, Zhejiang University, China}
\affiliation{Department of Physics, The University of Hong Kong, Hong Kong SAR, China}
\affiliation{Collaborative Innovation Center of Advanced Microstructures, Nanjing University, Nanjing, 210093, China}

\begin{abstract}
The dynamics of magnetization coupled to an electron gas via s-d exchange interaction is investigated by using density matrix technique.   Our theory shows that non-equilibrium spin accumulation induces a spin torque and the electron bath leads to a damping of the magnetization. For the two-dimensional magnetization thin film coupled to the electron gas with Rashba spin-orbit coupling, the result for the spin-orbit torques is consistent with the previous semi-classical theory.  Our theory predicts a damping of the magnetization, which is absent in the semi-classical theory. The magnitude of the damping due to the electron bath is comparable to the intrinsic Gilbert damping and may be important in describing the magnetization dynamics of the system.
\end{abstract}
\maketitle
\section{Introduction}
In study of  spin transfer torque (STT), it has been proposed \cite{STT1,STT2} to manipulate magnetic order parameter dynamics by using non-equilibrium electron bath instead of external magnetic fields.  The proposal has already led to commercial products in spintronics engineering.  Recently, there has been much attention on the "spin-orbit torque"(SOT), which was first proposed in theory\cite{SOT1,SOT2}, and later confirmed in experiments\cite{Exp1,Exp2,Exp3,Exp4} (see Ref.~\onlinecite{SOTrev1,SOTrev2} for a comprehensive review). After applying an external electric field to the electron gas with spin-orbit interaction(SOI), a component of the accumulated electron spin density mis-aligned with the ferromagnetic ordering can be created\cite{SOT1,SOT2}, which then will induce a field-like torque. The SOT opens the possibility of manipulating the magnetic order parameter in collinear magnetic structures and may efficiently reduce the critical current density for magnetization switching\cite{SOT1,SOT2}.  In the theoretical side, a full quantum theory has been proposed and developed to describe the dynamics of a single domain magnet under the continuous scattering by spin-polarized electrons. The quantum STT theory recovers the results of the semiclassical STT theory, and has revealed more details about the magnetization dynamics in the STT\cite{YSham1,YSham2,TSham}.  Therefore, it will be natural to apply a full quantum theory to study the magnetization dynamics influenced by the SOI electron gas. This may be an extension of the quantum STT theory to SOT.  In the full quantum theory, the quantum dynamics of the magnetization can be described by the evolution of its density matrix under the influence of the electron gas, which can be tuned by the external electric field. This treatment will not only give the mean-field effect on the magnetization dynamics by the electron bath, but also include the damping of the magnetization due to the fluctuation of the electron spin. The similar strategy has been exploited to investigate the photo-excited dynamics of the order parameter in Peierls chain\cite{CDW}.

This paper is organized as follows. In section II, we apply density matrix technique to derive general formalism for the magnetization dynamics driven by the electron bath through s-d exchange interaction. In section III, we apply the general formalism to the special case where the spatially uniform magnetization is coupled to a two dimensional electron gas with Rashba SOI, and calculate the spin-orbit torque and the damping effect of the electron bath. The main results are summarized and discussed in section IV.

\begin{figure}[]
\includegraphics[scale=0.40,clip]{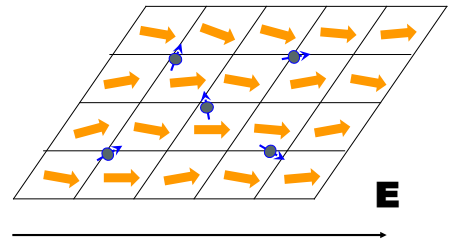}
\caption{(Color online). Schematic diagram for the lattice of localized spins (orange) coupled to the conductions electrons (blue) through s-d exchange interaction. An external electric field $\mathbf{E}$ can be applied to tune the electron bath. }\label{magnet}
\end{figure}
\section{General Formalism}
We apply density matrix technique to study dynamics of the magnetization driven by the electron bath via an s-d exchange interaction. The system is schematically illustrated in Fig.~\ref{magnet}, where the electron bath can be tuned by an external electric field. The Hamiltonian of the total system is formally written as
\begin{eqnarray}
H=H_{M}+H_{e}+H_{sd}.\label{Htot}
\end{eqnarray}
Here, $H_{M}$ is the Hamiltonian for the magnetization subsystem in terms of the local spin operators $\widehat{S}_{i,\mu}$ at site $i$ with spin directions $\mu(=x,y,z)$; $H_{e}$ is the Hamiltonian of the electron subsystem; $H_{sd}$ describes the s-d exchange interaction between the magnetization and the electron, where
\begin{eqnarray}
H_{sd}=J\sum_{i,\mu}\widehat{S}_{i,\mu}\widehat{\sigma}_{i,\mu}.\label{Hsd}
\end{eqnarray}
Here, $\widehat{\sigma}_{i,\mu}$ represents the electron spin operator at site $i$ without $\hbar/2$, and $J$ is the exchange coupling strength. Note that we have not specified the forms of $H_{M}$ and $H_{e}$ yet, thus the results below will be quite general.

The effect of the s-d exchange interaction $H_{sd}$ is twofold. On one hand, the magnetization dynamics is driven by the electron bath via $H_{sd}$; on the other hand, the electron states are also affected by the magnetization configuration in turn due to $H_{sd}$.  Since the time scale of the electron dynamics is usually much faster than that of the magnetization dynamics, we may assume that the electrons under the bias voltage establish a stationary non-equilibrium distribution in a very short time interval, during which the change of the magnetization configuration is negligible and the non-equilirium electron bath is approximated to be constant. The validity of this assumption only holds if the spin-lattice interaction is stronger than the s-d exchange interaction to relax the electron spin. Consider a short time interval $[t_{0},t]$, where the initial density matrices of the magnetization and the electron bath are $\rho_{M}(t_{0})$ and $\rho_{e}(t_{0})$ respectively. Then the initial magnetization configuration at each site is $S_{i,\mu}(t_{0})=\text{Tr}[\widehat{S}_{i,\mu}\rho_{M}(t_{0})]$, and the initial electron density matrix $\rho_{e}(t_{0})$ is determined by the bath Hamiltonian $H_{B}=H_{e}+J\sum_{i,\mu}S_{i,\mu}(t_{0})\widehat{\sigma}_{i,\mu}$ and the open boundary conditions.

In order to investigate the magnetization dynamics during the time interval $[t_{0},t]$ defined above, we redefine the local spin operators $\widehat{S}_{i,\mu}=S_{i,\mu}(t_{0})+\widehat{s}_{i,\mu}$, then the Hamiltonian $H$ in Eq.~(\ref{Htot}) can be rewritten as
\begin{eqnarray}
H=H_{M}+H_{B}+V_{sd},\label{Htot2}
\end{eqnarray}
with the interaction term
\begin{eqnarray}
V_{sd}=J\sum_{i,\mu}\widehat{s}_{i,\mu}\widehat{\sigma}_{i,\mu}.\label{Vsd}
\end{eqnarray}
During this time interval, the electron density matrix $\rho_{e}$ may be approximated to be constant because of the negligible change of the magnetization, and this can be justified in the limit $t\rightarrow t_{0}$. Assuming the total density matrix as $\rho(t)=\rho_{M}(t)\otimes\rho_{e}(t_0)$ and to the second order of interaction strength, the equation for the density matrix $\widetilde{\rho}_{M}(t)$ in the interaction picture is\cite{DenmaxBook}
\begin{eqnarray}
&&\frac{d}{dt}\widetilde{\rho}_{M}(t)=\frac{J}{i\hbar}\sum_{i,\mu}\sigma_{i,\mu}(t)
[\widetilde{s}_{i,\mu}(t),\widetilde{\rho}_{M}(t_{0})]\nonumber\\
&+&(\frac{J}{i\hbar})^{2}\sum_{i,\mu;j,\nu}\int_{t_{0}}^{t}d\tau\{\mathcal{C}_{i,\mu;j,\nu}(t,\tau)[\widetilde{s}_{i,\mu}(t),\widetilde{s}_{j,\nu}(\tau)\widetilde{\rho}_{M}(\tau)]\nonumber\\
&&-\mathcal{C}_{j,\nu;i,\mu}(\tau,t)[\widetilde{s}_{i,\mu}(t),\widetilde{\rho}_{M}(\tau)\widetilde{s}_{j,\nu}(\tau)]\}.\label{denmax}
\end{eqnarray}
Here, $\widetilde{\cdots}$ denotes the operators in the interaction picture; the electron spin polarization is $\sigma_{i,\mu}(t)=\text{Tr}_{e}[\widetilde{\sigma}_{i,\mu}(t)\widetilde{\rho}_{e}(t_{0})]$; the electron spin-spin correlation function is $\mathcal{C}_{i,\mu;j,\nu}(t,\tau)=\text{Tr}_{e}[\widetilde{\sigma}_{i,\mu}(t)\widetilde{\sigma}_{j,\nu}(\tau)\widetilde{\rho}_{e}(t_{0})]$, which is a function of $t-\tau$ only and satisfies the relation $\mathcal{C}_{i,\mu;j,\nu}(t,\tau)=\mathcal{C}_{j,\nu;i,\mu}^{*}(\tau,t)$. In priciple, the solution of Eq.~(\ref{denmax}) gives the density matrix of the magnetization in the time interval $[t_{0},t]$ under the influence of the electron bath, and can be applied to study the physical qualities that we are particularly interested in.

Based on Eq.~(\ref{denmax}), the dynamical equation for $S_{l,\lambda}(t)=\text{Tr}_{M}[\widetilde{S}_{l,\lambda}(t)\widetilde{\rho}_{M}(t)]$ is obtained as
\begin{eqnarray}
&&\frac{d}{dt}S_{l,\lambda}(t)=\frac{1}{i\hbar}\langle[\widehat{S}_{l,\lambda},H_{M}]\rangle_{t}+\frac{J}{\hbar}\sum_{\mu,\nu}\epsilon_{\lambda\mu\nu}\sigma_{l,\mu}(t)S_{l,\nu}(t)\nonumber\\&+&i(\frac{J}{i\hbar})^{2}\sum_{j,\mu,\nu,\xi}\epsilon_{\lambda\mu\nu}\int_{t_{0}}^{t}d\tau\{\mathcal{C}_{l,\mu;j,\xi}(t,\tau)\langle\widehat{S}_{l,\nu}(t)\widehat{s}_{j,\xi}(\tau)\rangle_{\tau}\nonumber\\
&&-\mathcal{C}_{j,\xi;l,\mu}(\tau,t)\langle\widehat{s}_{j,\xi}(\tau)\widehat{S}_{l,\nu}(t)\rangle_{\tau}\}.\label{mag}
\end{eqnarray}
Here, $\langle\cdots\rangle_{t}\equiv\text{Tr}_{M}[\cdots\rho_{M}(t)]$, and the spin commutation relation $[\widehat{S}_{l,\lambda},\widehat{S}_{i,\mu}]=i\delta_{li}\sum_{\nu}\epsilon_{\lambda\mu\nu}\widehat{S}_{l,\nu}$ has been exploited. The first term in the \emph{r.h.s.} of Eq.~(\ref{mag}) gives the intrinsic magnetization dynamics due to $H_{M}$; the second term is the spin torque term due to the accumulation of the electron spin density; the third term gives the damping effect of the electron bath. If the operator $\widehat{S}_{l,\nu}(t)$ in the damping term is approximately replaced by its expectation value $S_{l,\nu}(t)$, Eq.~(\ref{mag}) becomes
\begin{eqnarray}
&&\frac{d}{dt}S_{l,\lambda}(t)=\frac{1}{i\hbar}\langle[\widehat{S}_{l,\lambda},H_{M}]\rangle_{t}+\frac{J}{\hbar}\sum_{\mu,\nu}\epsilon_{\lambda\mu\nu}\sigma_{l,\mu}(t)S_{l,\nu}(t)\nonumber\\
&&+\frac{2J^{2}}{\hbar^{2}}\sum_{j,\mu,\nu,\xi}\epsilon_{\lambda\mu\nu}S_{l,\nu}(t)\int_{t_{0}}^{t}d\tau
\mathcal{K}_{l,\mu;j,\xi}(t-\tau)s_{j,\xi}(\tau),\label{mag1}
\end{eqnarray}
where $\mathcal{K}_{l,\mu;j,\xi}(t-\tau)$ is the imaginary part of $\mathcal{C}_{l,\mu;j,\xi}(t,\tau)$, and $s_{j,\xi}(\tau)=\langle\widehat{s}_{j,\xi}\rangle_{\tau}$. We introduce the kernel function $\gamma(t)$ which satisfies the relation $d\gamma_{l,\mu;j,\xi}(t)/dt=\mathcal{K}_{l,\mu;j,\xi}(t)$. The integral in the last term in Eq.~(\ref{mag1}) is rewritten as $\int_{t_{0}}^{t}d\tau\gamma_{l,\mu;j,\xi}(t-\tau)\dot{S}_{j,\xi}(\tau)$ after integrating by parts and neglecting the boundary terms in the limiting case $t\rightarrow t_{0}$. It can be further simplified as $\Gamma_{l,\mu;j,\xi}\dot{S}_{j,\xi}$ under the Markovian approximation $\dot{S}_{j,\xi}(\tau)\approx\dot{S}_{j,\xi}(t)$, with the coefficient $\Gamma_{l,\mu;j,\xi}=\int_{0}^{\delta t}d\tau\gamma_{l,\mu;j,\xi}(\tau)$ for $\delta t=t-t_{0}$.

Based on the discussions above, Eq.~(\ref{mag1}) can be written in a compact form
\begin{eqnarray}
\frac{d}{dt}\mathbf{S}_{l}(t)=\frac{1}{i\hbar}\langle[\widehat{\mathbf{S}}_{l},H_{M}]\rangle_{t}+\gamma_{e}\mathbf{B}_{l}(t)\times\mathbf{S}_{l}(t),\label{mag2}
\end{eqnarray}
where $\gamma_{e}$ is the gyromagnetic ratio; $\mathbf{B}_{l}$ is the effective magnetic field on the the local spin $\mathbf{S}_{l}$ originating from the electron bath. The $\mu$-component of $\mathbf{B}_{l}$ is expressed as
\begin{eqnarray}
B_{l,\mu}(t)=\frac{J}{\gamma_{e}\hbar}\sigma_{l,\mu}(t)+\frac{2J^{2}}{\gamma_{e}\hbar^{2}}\sum_{j,\xi}\Gamma_{l,\mu;j,\xi}(t)\dot{S}_{j,\xi}(t).\label{Bl}
\end{eqnarray}
The first term in (\ref{Bl}) will give the torque term due to the electron spin accumulation, which has been discussed extensively in previous studies; the second term will give the damping effect of the electron bath on the magnetization dynamics, which only emerges in the quantum treatment. The non-local feature of the damping term can be found here, which depends on the spatial correlation of $\Gamma_{l,\mu;j,\xi}$.

So far we have established a general dynamical equation for the magnetization when it is coupled to the electron bath via s-d exchange interaction. Here, both the Hamiltonian for the magnetization subsystem $H_{M}$ and the Hamiltonian for the electron subsystem $H_{e}$ have not been specified yet. The treatment is similar to the previous work on the order parameter dynamics in the photo-excited Peierls chain\cite{CDW}. In the next section, we apply this general formula to study the magnetization dynamics of a two-dimensional ferromagnetic thin film under the influence of an electron gas with Rashba SOI, i.e. a model system for ``spin-orbit torque".

\section{Spin-Orbit Torque}
\subsection{Electron Bath with Rashba SOI}
We consider a special system studied by Manchon and Zhang\cite{SOT1} for the spin-orbit torque. The two-dimensional magnetization thin film in x-y plane consists of $\mathcal{N}=M\times N$ lattice sites with the lattice constant $a$, and we will use the discrete notations in both real and reciprocal space. The magnetization is assumed to be uniform due to strong exchange interaction. The lack of inversion symmetry in $z$-direction induces the Rashba spin-orbit interaction in the two-dimensional electron gas. In this case, the Hamiltonian for the electron bath is given as\cite{SOT1}
\begin{eqnarray}
H_{B}=\frac{\widehat{\mathbf{p}}^{2}}{2m_{e}^{*}}+\frac{\alpha_{R}}{\hbar}(\widehat{\mathbf{p}}\times\widehat{\mathbf{z}})\cdot\widehat{\bm{\sigma}}+J\mathbf{S}\cdot\widehat{\bm{\sigma}},\label{RashbaH}
\end{eqnarray}
where $\widehat{\mathbf{p}}$ is the electron momentum operator; $m_{e}^{*}$ is the effective mass of electrons; $\alpha_{R}$ is the Rashba interaction strength; $\mathbf{S}=\mathbf{S}_{i}$ is the localized spin at each site. For $\mathbf{S}=S(\sin\theta\cos\phi,\sin\theta\sin\phi,\cos\theta)$, the energy dispersion relation of the electron is
\begin{eqnarray}
E_{\mathbf{k},\pm}=\frac{\hbar^{2}k^{2}}{2m_{e}^{*}}\pm\Delta_{\mathbf{k}}.\label{Ek}
\end{eqnarray}
Here, we have denoted the electron wavevector $\mathbf{k}=k(\cos\varphi,\sin\varphi)$, and
\begin{eqnarray*}
\Delta_{\mathbf{k}}=\sqrt{J^{2}S^{2}+\alpha_{R}^{2}k^{2}-2JS\alpha_{R}k\sin\theta\sin(\phi-\varphi)}.
\end{eqnarray*}
The corresponding electron eigenstates $|\mathbf{k},\pm\rangle$ are
\begin{eqnarray}
|\mathbf{k},\pm\rangle=\frac{1}{\sqrt{\mathcal{N}}}e^{i\mathbf{k}\cdot\mathbf{r}}\left(\begin{array}{c}\cos\frac{\Theta_{\mathbf{k},\pm}}{2}e^{-i\Phi_{\mathbf{k}}}\\\sin\frac{\Theta_{\mathbf{k},\pm}}{2}\end{array}\right),\label{veck}
\end{eqnarray}
where the angles $\Theta_{\mathbf{k},\pm}$ and $\Phi_{\mathbf{k}}$ are determined by
\begin{eqnarray*}
\cos\frac{\Theta_{\mathbf{k},\pm}}{2}&=&\frac{\sqrt{\Delta_{\mathbf{k}}^{2}-J^{2}S^{2}\cos^{2}\theta}}{\sqrt{2\Delta_{\mathbf{k}}^{2}\mp 2JS\Delta_{\mathbf{k}}\cos\theta}},\\
\sin\frac{\Theta_{\mathbf{k},\pm}}{2}&=&\frac{\pm\Delta_{\mathbf{k}}-JS\cos\theta}{\sqrt{2\Delta_{\mathbf{k}}^{2}\mp 2JS\Delta_{\mathbf{k}}\cos\theta}},\\
\cos\Phi_{\mathbf{k}}&=&\frac{JS\sin\theta\cos\phi+\alpha_{R}k\sin\varphi}{\sqrt{\Delta_{\mathbf{k}}^{2}-J^{2}S^{2}\cos^{2}\theta}},\\
\sin\Phi_{\mathbf{k}}&=&\frac{JS\sin\theta\sin\phi-\alpha_{R}k\cos\varphi}{\sqrt{\Delta_{\mathbf{k}}^{2}-J^{2}S^{2}\cos^{2}\theta}}.
\end{eqnarray*}
The spin polarization vector for the state $|\mathbf{k},\pm\rangle$ is $\mathbf{P}_{\mathbf{k},\pm}=(\sin\Theta_{\mathbf{k},\pm}\cos\Phi_{\mathbf{k}},\sin\Theta_{\mathbf{k},\pm}\sin\Phi_{\mathbf{k}},\cos\Phi_{\mathbf{k},\pm})$.

The statistical properties of the electron bath are determined by the probability distribution function $f_{\mathbf{k},s}$ for the state $|\mathbf{k},s=\pm\rangle$, which can be tuned by the external field. If an electric field $\mathbf{E}$ is applied, the non-equilibrium distribution of the electron states will be established due to the random scattering potential by impurities\cite{SOT1}. The distribution function $f_{\mathbf{k},s}$ is determined by the Boltzmann equation
\begin{eqnarray}
-\frac{e\mathbf{E}}{\hbar}\cdot\nabla_{\mathbf{k}}f_{\mathbf{k},s}=\mathcal{S}_{c}[f_{\mathbf{k},s}].\label{Boltz}
\end{eqnarray}
The collision integral $\mathcal{S}_{c}[f_{\mathbf{k},s}]$ describes the relaxation of the occupied state $|
\mathbf{k},s\rangle$ and can be treated by the relaxation time approximation, namely,
\begin{eqnarray}
\mathcal{S}_{c}[f_{\mathbf{k},s}]=-\frac{f_{\mathbf{k},s}-f_{\mathbf{k},s}^{0}}{\tau}.\label{col}
\end{eqnarray}
Here, $f_{\mathbf{k},s}^{0}$ is the equilibrium distribution function, and an isotropic relaxation time $\tau$ has been assumed\cite{SOT1}. To the first order of the electric field, the solution of Eq.~(\ref{Boltz}) is $f_{\mathbf{k},s}=f_{\mathbf{k},s}^{0}+g_{\mathbf{k},s}$, where the out of equilibrium part induced by the external electric field is
\begin{eqnarray}
g_{\mathbf{k},s}=\frac{\partial f_{\mathbf{k},s}^{0}}{\partial E}e\mathbf{E}\cdot\mathbf{v}_{\mathbf{k},s}\tau,\label{dfks}
\end{eqnarray}
with the electron velocity $\mathbf{v}_{\mathbf{k},s}=\frac{1}{\hbar}\nabla_{\mathbf{k}}E_{\mathbf{k},s}$. Such a treatment of the non-equilirium electron distribution was also exploited in the previous semiclassical theory\cite{SOT1}.

\subsection{Electron Spin Polarization and Torque}
With the non-equilibrium distribution function $f_{\mathbf{k},s}$ given above, the electron spin polarization $\sigma_{l,\mu}$ at site $l$ and the correlation function $\mathcal{C}_{l,\mu;j,\xi}(t,\tau)$ in Eq.~(\ref{Bl}) can be calculated, and the torque and damping effect due to the electron bath can be obtained. In the second quantization representation of the basis set $\{|\mathbf{k},s\rangle\}$, the operator $\widehat{\sigma}_{l,\mu}$ is expressed as
\begin{eqnarray}
\widehat{\sigma}_{l,\mu}=\frac{1}{\mathcal{N}}\sum_{\mathbf{k},s;\mathbf{k}',s'}\chi_{\mathbf{k},s;\mathbf{k}',s'}^{\mu}e^{i(\mathbf{k}'-\mathbf{k})\cdot\mathbf{r}_{l}}\widehat{c}_{\mathbf{k},s}^{\dag}\widehat{c}_{\mathbf{k}',s'},\nonumber
\end{eqnarray}
where the matrix element
\begin{eqnarray}
\chi_{\mathbf{k},s;\mathbf{k}',s'}^{\mu}=(\cos\frac{\Theta_{\mathbf{k},s}}{2}e^{i\Phi_{\mathbf{k}}},\sin\frac{\Theta_{\mathbf{k},s}}{2})\sigma_{\mu}
\left(\begin{array}{c}\cos\frac{\Theta_{\mathbf{k}',s'}}{2}e^{-i\Phi_{\mathbf{k}'}}\\\sin\frac{\Theta_{\mathbf{k}',s'}}{2}\end{array}\right).\nonumber
\end{eqnarray}
Then the electron spin polarization $\sigma_{l,\mu}$ is
\begin{eqnarray}
\sigma_{l,\mu}=\frac{1}{\mathcal{N}}\sum_{\mathbf{k},s}\chi_{\mathbf{k},s;\mathbf{k},s}^{\mu}f_{\mathbf{k},s}=\frac{1}{\mathcal{N}}\sum_{\mathbf{k},s}P_{\mathbf{k},s}^{\mu}f_{\mathbf{k},s}.\label{spinden}
\end{eqnarray}
For the physically relevant case $\alpha_{R}k\ll JS$, the approximate value of $\mathbf{P}_{\mathbf{k},\pm}$ to the first order of $\frac{\alpha_{R}k}{JS}$ is
\begin{eqnarray}
\mathbf{P}_{\mathbf{k},\pm}&=&\pm\left(\begin{array}{c}\mathsf{S}_{x}+\frac{\alpha_{R}}{JS}\mathsf{S}_{x}\mathsf{S}_{y} k_{x}+\frac{\alpha_{R}}{JS}(1-\mathsf{S}_{x}^{2})k_{y}\\\mathsf{S}_{y}-\frac{\alpha_{R}}{JS}(1-\mathsf{S}_{y}^{2})k_{x}-\frac{\alpha_{R}}{JS}\mathsf{S}_{x}\mathsf{S}_{y}k_{y}\\\mathsf{S}_{z}+\frac{\alpha_{R}}{JS}\mathsf{S}_{y}\mathsf{S}_{z}k_{x}-\frac{\alpha_{R}}{JS}\mathsf{S}_{x}\mathsf{S}_{z}k_{y}\end{array}\right).\nonumber
\end{eqnarray}
Here, the unit vector for the magnetization is denoted as $\mathbf{\mathsf{S}}=(\sin\theta\cos\phi,\sin\theta\sin\phi,\cos\phi)$.

For the electric current density $\mathbf{j}_{e}=j_{e}(\cos\vartheta,\sin\vartheta,0)$, the non-equilibrium spin polarization $\delta\bm\sigma_{l}$ which is perpendicular to $\mathbf{\mathsf{S}}$ is calculated to be (Appendix A)
\begin{eqnarray*}
\delta\bm\sigma_{l}=-\frac{\alpha_{R}m_{e}^{*}j_{e}a^{3}}{e\hbar E_{f}}\left(\begin{array}{c}\cos\vartheta\mathsf{S}_{x}\mathsf{S}_{y}+\sin\vartheta(1-\mathsf{S}_{x}^{2})\\-\cos\vartheta(1-\mathsf{S}_{y}^{2})-\sin\vartheta\mathsf{S}_{x}\mathsf{S}_{y}\\
\cos\vartheta\mathsf{S}_{y}\mathsf{S}_{z}-\sin\vartheta\mathsf{S}_{x}\mathsf{S}_{z}\end{array}\right),
\end{eqnarray*}
where $E_{f}$ denotes the Fermi energy. The torque $\mathbf{T}_{l}$ is then obtained as
\begin{eqnarray*}
\mathbf{T}_{l}&=&\frac{JS\alpha_{R}m_{e}^{*}j_{e}a^{3}}{e\hbar^{2}E_{f}}\left(\begin{array}{c}\cos\vartheta\mathsf{S}_{z}\\\sin\vartheta\mathsf{S}_{z}\\-\cos\vartheta\mathsf{S}_{x}-\sin\vartheta\mathsf{S}_{y}\end{array}\right)\nonumber\\
&=&\frac{J\alpha_{R}m_{e}^{*}a^{3}}{e\hbar^{2}E_{f}}(\widehat{\mathbf{z}}\times\mathbf{j}_{e})\times\mathbf{S}_{l}.
\end{eqnarray*}
This result reproduces the form of SOT obtained before\cite{SOT1}, but the magnetization vector is not restricted in two-dimensional x-y plane in our derivations. It is easily understood from the effective Hamiltonian (\ref{RashbaH}), where the non-equilibrium distribution of electron states will produce an extra electron spin polarization along the direction $\widehat{\mathbf{z}}\times\mathbf{j}_{e}$.

\subsection{Correlation Function and Damping}
We now calculate the correlation function $\mathcal{C}_{l,\mu;j,\xi}(t,\tau)$,  which gives the damping term for the magnetization dynamics due to the electron bath. Since $\widehat{c}_{\mathbf{k},s}(t)=\widehat{c}_{\mathbf{k},s}e^{-iE_{\mathbf{k},s}t/\hbar}$, the correlation function $\mathcal{C}_{l,\mu;j,\xi}(t,\tau)$ is formally written as
\begin{eqnarray}
&&\mathcal{C}_{l,\mu;j,\xi}(t,\tau)\nonumber\\&=&\frac{1}{\mathcal{N}^{2}}\sum_{\mathbf{k},s;\mathbf{k}',s'}\sum_{\mathbf{k}'',s'';\mathbf{k}''',s'''}e^{i(\mathbf{k}'-\mathbf{k})\cdot\mathbf{r}_{l}}e^{i(\mathbf{k}'''-\mathbf{k}'')\cdot\mathbf{r}_{j}}
\nonumber\\&\times&e^{i(E_{\mathbf{k},s}-E_{\mathbf{k}',s'})t/\hbar}e^{i(E_{\mathbf{k}'',s''}-E_{\mathbf{k}''',s'''})\tau/\hbar}\nonumber\\
&\times&\chi_{\mathbf{k},s;\mathbf{k}',s'}^{\mu}\chi_{\mathbf{k}'',s'';\mathbf{k}''',s'''}^{\xi}\langle\widehat{c}_{\mathbf{k},s}^{\dag}\widehat{c}_{\mathbf{k}',s'}
\widehat{c}_{\mathbf{k}'',s''}^{\dag}\widehat{c}_{\mathbf{k}''',s'''}\rangle.
\end{eqnarray}
We see that $\mathcal{C}_{l,\mu;j,\xi}(t,\tau)$ is the function of $\mathbf{r}_{l}-\mathbf{r}_{j}$ and $t-\tau$, due to the space and time translation invariance for the investigated system. For simplicity, we estimate $\mathcal{C}_{l,\mu;j,\xi}(t,\tau)$ with several approximations. Firstly, we assume that the phase factor $e^{i(\mathbf{k}'-\mathbf{k})\cdot(\mathbf{r}_{l}-\mathbf{r}_{j})}$ will cause the cancellation of the summations over $\mathbf{k}$ and $\mathbf{k}'$ if $\mathbf{r}_{l}\neq\mathbf{r}_{j}$, thus $\mathcal{C}_{l,\mu;j,\xi}=\mathcal{C}_{\mu\xi}\delta_{lj}$. Secondly, $\chi_{\mathbf{k},s;\mathbf{k}',s'}^{\mu}$ are calculated to the zeroth order of $\frac{\alpha_{R}k}{JS}$ for the relevant case $\alpha_{R}k\ll JS$, where the electron spin states are $\mathbf{k}$-independent, i.e.
\begin{eqnarray}
\chi_{\pm\pm}&=&\pm(\sin\theta\cos\phi,\sin\theta\sin\phi,\cos\theta),\nonumber\\
\chi_{+-}&=&(-\cos\theta\cos\phi-i\sin\phi,-\cos\theta\sin\phi+i\cos\phi,\sin\theta).\nonumber
\end{eqnarray}
Furthermore, we calculate the correlation function $\langle\widehat{c}_{\mathbf{k},s}^{\dag}\widehat{c}_{\mathbf{k}',s'}
\widehat{c}_{\mathbf{k}'',s''}^{\dag}\widehat{c}_{\mathbf{k}''',s'''}\rangle$ with the electron bath at equilibrium, where the effect of the non-equilibrium electric current induced by the external field will be neglected. This enable us to apply the Wick contraction\cite{Wick} to simplify the calculations. The negligence of the dependence of the damping coefficient on the Rashba SOI and the non-equilibrium electric current is valid if the dynamical equation (\ref{mag2}) is kept to the first order of these two factors. With the above approximations, we get
\begin{eqnarray*}
&&\mathcal{C}_{\mu\xi}(t)\nonumber\\
&=&\frac{1}{\mathcal{N}^{2}}\sum_{\mathbf{k},s}\chi_{ss}^{\mu}\chi_{ss}^{\xi}f_{\mathbf{k},s}+\frac{1}{\mathcal{N}^{2}}\sum_{\mathbf{k},s;\mathbf{k}',s'}\chi_{ss}^{\mu}\chi_{s's'}^{\xi}f_{\mathbf{k},s}f_{\mathbf{k}',s'}
\nonumber\\&+&\frac{1}{\mathcal{N}^{2}}\sum_{\mathbf{k},s;\mathbf{k}',s'}e^{i(E_{\mathbf{k},s}-E_{\mathbf{k}',s'})t/\hbar}\chi_{ss'}^{\mu}\chi_{s's}^{\xi}f_{\mathbf{k},s}(1-f_{\mathbf{k}',s'}),
\end{eqnarray*}
where $|\mathbf{k},s\rangle$ and $|\mathbf{k}',s'\rangle$ are different states.

\begin{figure}[htbp]
\includegraphics[scale=0.26,clip]{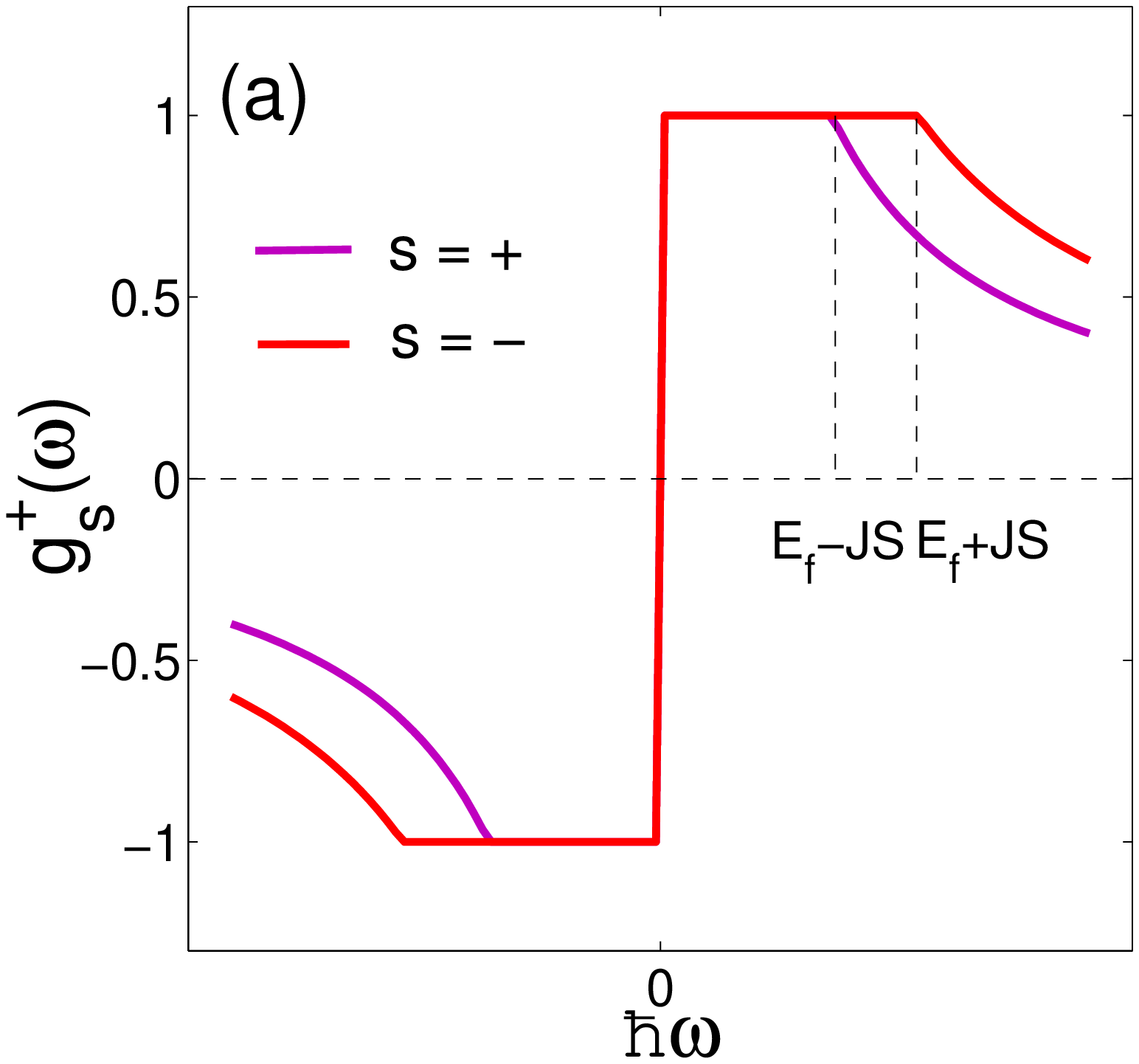}
\includegraphics[scale=0.26,clip]{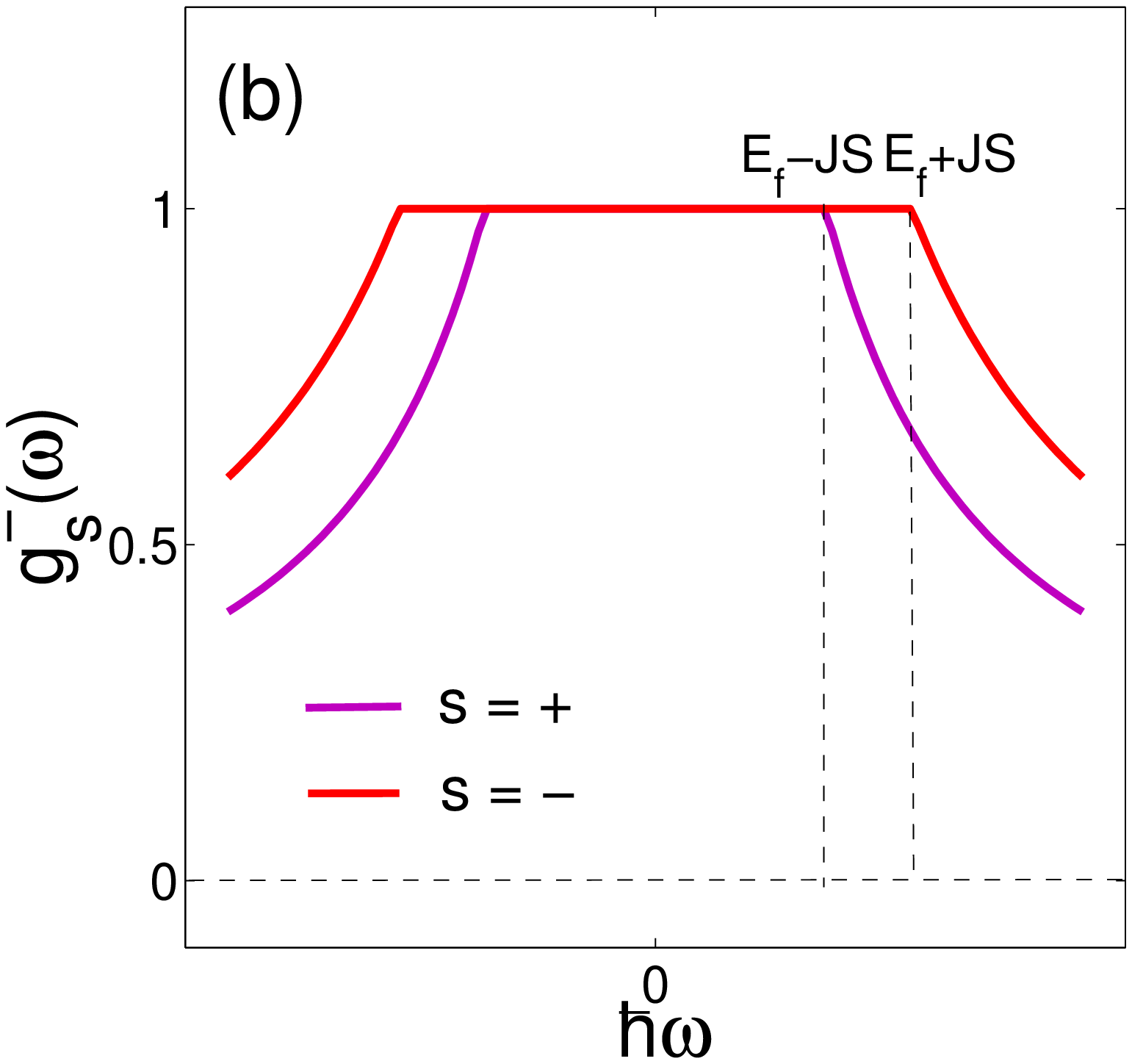}
\caption{(Color online). Schematic diagram for $g_{s}^{\pm}(\omega)$. Blue line for $s=+$, and red line for $s=-$. Notice that $g_{s}^{+}$ is an odd function of $\omega$ and $g_{s}^{-}(\omega)$ is an even function of $\omega$, and they approach to $0$ when $|\omega|\rightarrow \infty$.  }\label{gomg}
\end{figure}

Since the kernel function $\gamma_{l,\mu;j,\xi}(t)$ is given by the relation $d\gamma_{l,\mu;j,\xi}(t)/dt=\mathcal{K}_{l,\mu;j,\xi}(t)$, where $\mathcal{K}_{l,\mu;j,\xi}(t)=\Im[\mathcal{C}_{l,\mu;j,\xi}(t)]$, their Fourier transformations are related by $\gamma_{l,\mu;j,\xi}(\omega)=\frac{i}{\omega}\mathcal{K}_{l,\mu;j,\xi}(\omega)$. The Fourier transformation of $\mathcal{K}_{l,\mu;j,\xi}(t)$ is obtained as (Appendix B)
\begin{eqnarray*}
&&\mathcal{K}_{l,\mu;j,\xi}(\omega)\nonumber\\&=&\delta_{lj}(\frac{m_{e}^{*}a^{2}}{2\pi\hbar^{2}})^{2}\frac{\hbar}{2i}\sum_{s,s'}[\chi_{ss'}^{\mu}\chi_{s's}^{\xi}g_{s}(\omega)-(\chi_{ss'}^{\mu}\chi_{s's}^{\xi})^{*}g_{s}(-\omega)],\nonumber
\end{eqnarray*}
where the function $g_{s}(\omega)$ is defined as
\begin{eqnarray*}
g_{s}(\omega)=\left\{\begin{array}{ccc}0&,&\hbar\omega<0;\\\hbar\omega&,&0<\hbar\omega<E_{f}-sJS;\\E_{f}-sJS&,&\hbar\omega>E_{f}-sJS.\end{array}\right.
\end{eqnarray*}
Then the damping kernel function $\gamma_{l,\mu;j,\xi}(t)$ can be calculated by the inverse Fourier transformation from $\gamma_{l,\mu;j,\xi}(\omega)$, which results in (Appendix B)
\begin{eqnarray}
\gamma_{l,\mu;j,\xi}(t)
=\delta_{lj}(\frac{m_{e}^{*}a^{2}}{2\pi\hbar})^{2}\frac{1}{2}\sum_{s}(\delta_{\mu\xi}g_{s}^{-}(t)+is\sum_{\nu}\epsilon_{\mu\xi\nu}\mathsf{S}_{\nu}g_{s}^{+}(t)).\nonumber\\\label{Gamt}
\end{eqnarray}
Here, $g_{s}^{\pm}(t)=\int_{-\infty}^{+\infty}d\omega g_{s}^{\pm}(\omega)e^{-i\omega t}$ and
$g_{s}^{\pm}(\omega)=\frac{1}{\hbar\omega}(g_{s}(\omega)\pm g_{s}(-\omega))$, as schematically shown in Fig.~\ref{gomg}. Then the coefficient $\Gamma_{l,\mu;j,\xi}$ in Eq.~(\ref{Bl}) is obtained as
\begin{eqnarray}
\Gamma_{l,\mu;j,\xi}=\delta_{lj}(\frac{m_{e}^{*}a^{2}}{2\pi\hbar})^{2}(\Gamma^{(1)}\delta_{\mu\xi}+\Gamma^{(2)}\sum_{\nu}\epsilon_{\mu\xi\nu}\mathsf{S}_{\nu}),\label{Gamma}
\end{eqnarray}
with $\Gamma^{(1)}=\frac{1}{2}\sum_{s}\int_{0}^{\delta t}d\tau g_{s}^{-}(\tau)$ and $\Gamma^{(2)}=\frac{i}{2}\sum_{s}s\int_{0}^{\delta t}d\tau g_{s}^{+}(\tau)$. Then the damping part in Eq.~(\ref{mag2}) can be explicitly written as
\begin{eqnarray}
\mathbf{D}_{l}=2(\frac{Jm_{e}^{*}a^{2}}{2\pi\hbar^{2}})^{2}(\Gamma^{(1)}\dot{\mathbf{S}}_{l}\times\mathbf{S}_{l}+S\Gamma^{(2)}\dot{\mathbf{S}}_{l}),\label{Dterm}
\end{eqnarray}
which is independent of the Rashba constant and the electric current due to our approximations above.

The first term in (\ref{Dterm}) will give the damping effect which drives the local spin towards the direction with the lower energy; while the second term in (\ref{Dterm}) will give a renormalized factor in Eq.~(\ref{mag2}). Assuming that $J\sim 1$~eV, $m_{e}^{*}\sim m_{e}$, $a\sim 1$~\AA, one gets the rough estimation of the magnitude order for the factor $(\frac{Jm_{e}^{*}a^{2}}{2\pi\hbar^{2}})^{2}\sim 10^{-3}$, thus the damping effect due to the electron bath is comparable to the intrinsic Gilbert damping of some ferromagnetic materials. This damping effect can become important to understand the dissipative features of the magnetization dynamics driven by spin-orbit torque.

\section{Conclusion}
In conclusion, we have applied density matrix technique to formulate the magnetization dynamics of a system consisting of local magnetic moments influenced by an electron gas through s-d exchange interaction. In this approach, the magnetic subsystem is treated as an open quantum system and the electron gas acts as a non-equilibrium bath tuned by the external electric field. The spin torque due to the non-equilibrium electron spin accumulation and the damping effect of the electron bath have been taken into account simultaneously. We apply the developed formula to the model system for spin-orbit torque, where the two-dimensional magnetization film is coupled to the Rashba electron gas through s-d exchange interaction. We have calculated the spin-orbit torque and the results are consistent with the previous study.  However, our method does not require the magnetization direction to be in the two-dimensional plane as in the previous study.  Our approach enables us to obtain the damping effect due to the electron bath, which is a new feature absent in the semiclassical theory. The damping caused by the electron bath is estimated to be comparable to the intrinsic Gilbert damping, and may be important to describe the magnetization dynamics driven by spin-orbit torque. In brief, this work has extended the previous semiclassical theory for spin-orbit torque to a more complete description. Further applications of this approach are expected to understand and to manipulate the magnetization dynamics through electron gas in other complex cases.

\begin{acknowledgements}
This work was supported in part by the Hong Kong's University Grant Council via grant AoE/P-04/08.  This work is also partially supported by National Basic Research Program of China (No. 2014CB921203), NSFC grant (No.11274269), and NSFC grant (No.11204186).
\end{acknowledgements}

\appendix
\section{Electron Spin Polarization}
We first assume that the electric field is applied along $x$-direction, then
\begin{eqnarray*}
\delta\bm\sigma_{l}=\frac{1}{\mathcal{N}}\sum_{\mathbf{k},s}g_{\mathbf{k},s}\mathbf{P}_{\mathbf{k},s}
=\frac{1}{\mathcal{N}}\sum_{\mathbf{k}}(g_{\mathbf{k},+}-g_{\mathbf{k},-})k_{x}\frac{\alpha_{R}}{JS}\bm\Sigma_{x},\nonumber
\end{eqnarray*}
where $\bm\Sigma_{x}=(\mathsf{S}_{x}\mathsf{S}_{y},-(1-\mathsf{S}_{y}^{2}),\mathsf{S}_{y}\mathsf{S}_{z})$.
The corresponding electric current density is
\begin{eqnarray*}
j_{e}=-\frac{e}{\mathcal{N}a^{3}}\sum_{\mathbf{k},s}g_{\mathbf{k},s}(v_{\mathbf{k},s})_{x}\approx -\frac{e\hbar}{m_{e}^{*}}\frac{1}{\mathcal{N}}\sum_{\mathbf{k},s}g_{\mathbf{k},s} k_{x},
\end{eqnarray*}
and the spin current density is
\begin{eqnarray*}
\mathbf{j}_{s}&=&\frac{\hbar}{2\mathcal{N}a^{3}}\sum_{\mathbf{k},s}g_{\mathbf{k},s}(v_{\mathbf{k},s})_{x}\mathbf{P}_{k,s}\nonumber\\&\approx&\frac{\hbar^{2}}{2m_{e}^{*}}\frac{1}{\mathcal{N}a^{3}}\sum_{\mathbf{k}}(g_{\mathbf{k},+}-g_{\mathbf{k},-})k_{x}\mathbf{\mathsf{S}}.
\end{eqnarray*}
Thus a rough relation is obtained as
\begin{eqnarray*}
\delta\bm\sigma_{l}=-\frac{\alpha_{R}m_{e}^{*}j_{e}a^{3}}{e\hbar E_{f}}\mathbf{\Sigma}_{x},
\end{eqnarray*}
where the relation $\mathbf{j}_{s}\approx -\frac{\hbar JS}{2eE_{f}}j_{e}\mathbf{\mathsf{S}}$ has been used here.

Similarly, if the electric field is applied along the $y$-direction, the non-equilibrium spin polarization will be
\begin{eqnarray*}
\delta\bm\sigma_{l}=-\frac{\alpha_{R}m_{e}^{*}j_{e}a^{3}}{e\hbar E_{f}}\mathbf{\Sigma}_{y},
\end{eqnarray*}
with
$\mathbf{\Sigma}_{y}=(1-\mathsf{S}_{x}^{2},-\mathsf{S}_{x}\mathsf{S}_{y},-\mathsf{S}_{x}\mathsf{S}_{z})$.
Therefore, for the electric current density $\mathbf{j}_{e}=j_{e}(\cos\vartheta,\sin\vartheta,0)$, we get
\begin{eqnarray*}
\delta\bm\sigma_{l}=-\frac{\alpha_{R}m_{e}^{*}j_{e}a^{3}}{e\hbar E_{f}}\left(\begin{array}{c}\cos\vartheta\mathsf{S}_{x}\mathsf{S}_{y}+\sin\vartheta(1-\mathsf{S}_{x}^{2})\\-\cos\vartheta(1-\mathsf{S}_{y}^{2})-\sin\vartheta\mathsf{S}_{x}\mathsf{S}_{y}\\
\cos\vartheta\mathsf{S}_{y}\mathsf{S}_{z}-\sin\vartheta\mathsf{S}_{x}\mathsf{S}_{z}\end{array}\right).
\end{eqnarray*}

\section{Correlation Function and Damping Kernel}
The imaginary part of $\mathcal{C}_{\mu\xi}(t)$ is given as
\begin{eqnarray*}
&&\mathcal{K}_{\mu\xi}(t)\\&=&\Im{[\mathcal{C}_{\mu\xi}(t)]}\nonumber\\&=&(\frac{m_{e}^{*}a^{2}}{2\pi\hbar^{2}})^{2}\sum_{s,s'}\Im[\chi_{ss'}^{\mu}\chi_{s's}^{\xi}\int_{sJS}^{E_{f}}d\epsilon\int_{E_{f}}^{\infty}d\epsilon'e^{\frac{i}{\hbar}(\epsilon-\epsilon')t}]\nonumber\\
&=&(\frac{m_{e}^{*}a^{2}}{2\pi\hbar^{2}})^{2}\sum_{s,s'}\int_{sJS}^{E_{f}}d\epsilon\int_{E_{f}}^{\infty}d\epsilon'[-\frac{i}{2}\chi_{ss'}^{\mu}\chi_{s's}^{\xi}e^{\frac{i}{\hbar}(\epsilon-\epsilon')t}+h.c.].
\end{eqnarray*}
Here, $f_{\mathbf{k},s}$ is approximated as the zero-temperature Fermi distribution function, and the relation $\frac{1}{\mathcal{N}}\sum_{\mathbf{k}}\rightarrow\frac{a^{2}}{(2\pi)^{2}}\int d^{2}\mathbf{k}=\frac{m_{e}^{*}a^{2}}{2\pi\hbar^{2}}\int d\epsilon$ has been used. Its Fourier transformation $\mathcal{K}_{\mu\xi}(\omega)$ is then
\begin{eqnarray*}
&&\mathcal{K}_{\mu\xi}(\omega)\nonumber\\&=&
\frac{1}{2\pi}\int_{-\infty}^{+\infty}dt\mathcal{K}_{\mu\xi}(t)e^{i\omega t}\nonumber\\
&=&(\frac{m_{e}^{*}a^{2}}{2\pi\hbar^{2}})^{2}\sum_{s,s'}\int_{sJS}^{E_{f}}d\epsilon\int_{E_{f}}^{\infty}d\epsilon'\\
&\times&[-\frac{i}{2}\chi_{ss'}^{\mu}\chi_{s's}^{\xi}\delta(\omega+\frac{\epsilon-\epsilon'}{\hbar})+\frac{i}{2}(\chi_{ss'}^{\mu}\xi_{s's}^{\xi})^{*}\delta(\omega+\frac{\epsilon'-\epsilon}{\hbar})]\nonumber\\
&=&-(\frac{m_{e}^{*}a^{2}}{2\pi\hbar^{2}})^{2}\frac{i\hbar}{2}\sum_{s,s'}[\chi_{ss'}^{\mu}\chi_{s's}^{\xi}g_{s}(\omega)-(\chi_{ss'}^{\mu}\chi_{s's}^{\xi})^{*}g_{s}(-\omega)],\nonumber
\end{eqnarray*}
where the function $g(\omega)$ is defined as
\begin{eqnarray*}
g_{s}(\omega)=\left\{\begin{array}{ccc}0&,&\hbar\omega<0;\\\hbar\omega&,&0<\hbar\omega<E_{f}-sJS;\\E_{f}-sJS&,&\hbar\omega>E_{f}-sJS.\end{array}\right.
\end{eqnarray*}
Therefore,
\begin{eqnarray*}
&&\gamma_{l,\mu;j,\xi}(\omega)\nonumber\\&=&\delta_{lj}(\frac{m_{e}^{*}a^{2}}{2\pi\hbar})^{2}\frac{1}{2}\sum_{s,s'}[\Re(\chi_{ss'}^{\mu}\chi_{s's}^{\xi})g_{s}^{-}(\omega)+i\Im(\chi_{ss'}^{\mu}\chi_{s's}^{\xi})g_{s}^{+}(\omega)],
\end{eqnarray*}
where $g_{s}^{\pm}(\omega)=\frac{1}{\hbar\omega}(g_{s}(\omega)\pm g_{s}(-\omega))$, and $\gamma_{l,\mu;j,\xi}(t)$ is calculated as
\begin{eqnarray*}
&&\gamma_{l,\mu;j,\xi}(t)\nonumber\\&=&\int_{-\infty}^{+\infty}d\omega\gamma_{l,\mu;j,\xi}(\omega)e^{-i\omega t}\nonumber\\
&=&\delta_{lj}(\frac{m_{e}^{*}a^{2}}{2\pi\hbar})^{2}\frac{1}{2}\sum_{s,s'}[\Re(\chi_{ss'}^{\mu}\chi_{s's}^{\xi})g_{s}^{-}(t)+i\Im(\chi_{ss'}^{\mu}\chi_{s's}^{\xi})g_{s}^{+}(t)]\nonumber\\
&=&\delta_{lj}(\frac{m_{e}^{*}a^{2}}{2\pi\hbar})^{2}\frac{1}{2}\sum_{s}(\delta_{\mu\xi}g_{s}^{-}(t)+is\sum_{\nu}\epsilon_{\mu\xi\nu}\mathsf{S}_{\nu}g_{s}^{+}(t))\nonumber\\
&\approx&\delta_{lj}\delta_{\mu\xi}(\frac{m_{e}^{*}a^{2}}{2\pi\hbar})^{2}g^{-}(t),\nonumber
\end{eqnarray*}
where $g_{s}^{\pm}(t)=\int_{-\infty}^{+\infty}d\omega g_{s}^{\pm}(\omega)e^{-i\omega t}$ and we have
used the expressions
\begin{eqnarray*}
\chi_{+,+}^{\mu}\chi_{+,+}^{\xi}&=&\chi_{-,-}^{\mu}\chi_{-,-}^{\xi}=\mathsf{S}_{\mu}\mathsf{S}_{\xi}.\nonumber\\
\chi_{+,-}^{\mu}\chi_{-,+}^{\xi}&=&(\chi_{-,+}^{\mu}\chi_{+,-}^{\xi})^{*}=\delta_{\mu\xi}-\mathsf{S}_{\mu}\mathsf{S}_{\xi}+i\sum_{\nu}\epsilon_{\mu\xi\nu}\mathsf{S}_{\nu}.
\end{eqnarray*}

\end{document}